# Why do we need social singularity? A mechanism-based critique of gradual scenarios in AI existential-risk discourse

Petr O. Jedlička[1]

*This paper critiques recent gradual and cumulative AI existential-risk scenarios, especially on the grounds that they remain insufficiently sociologically specified. While these scenarios make a substantive contribution, they lack a reflexive perspective, retain a largely technologically deterministic structure and underestimate the role of collective agency and other social processes. As a result, they also discount the possibility of major social conflict accompanying AI diffusion, which in turn limits their plausibility.*

*The paper further identifies a set of social mechanisms likely to become significant before or during AI deployment. These mechanisms include interpretive and performative dynamics; mobilisation and countermobilisation; path dependence and lock-in; cross-regime divergence and multi-speed diffusion; and social-psychological mechanisms such as attachment and reactance. These mechanisms will interact recursively with AI diffusion and reshape future trajectories, which may subsequently branch, reverse, or undergo discontinuous shifts – thereby expanding the space of plausible AI futures.*

*On this basis, the paper also proposes a novel analytic distinction between technological singularity and social singularity. Whereas the former refers to a putative technological threshold in AI development, the latter denotes a social discontinuity produced by the anticipated approach of that threshold. The central implication is that the most consequential disruptions may emerge not only from advanced AI itself, but also from the social dynamics provoked by the expectation of its arrival.*

## 1. Introduction

### 1.1. Gradual scenarios: their contributions and omissions

Recently, a new class of AI x-risk scenarios and models has emerged, putting forward a hypothesis of gradual disempowerment rather than sudden large-scale AI takeover events typical of conventional accounts of existential catastrophe in which AI systems rapidly seize control (Yudkowsky 2008; Yampolskiy 2015; Bostrom 2014; Carlsmith 2022; Ord 2020; Karnofsky 2023; Bengio et al. 2026; Kokotajlo et al. 2025). This new class of AI x-risk scenarios and models[2] treats advanced AI risk as unfolding through the incremental manifestations of disruptions caused by the emergence of advanced AI systems – thus creating an alternative pathway for AI futures.

[1] Institute of Philosophy, Czech Academy of Sciences, Jilská 361/1, 110 00 Prague 1, Czechia, jedlicka@flu.cas.cz

[2] In what follows, "gradual scenarios and models" serves as an umbrella term for this class of accounts.

The gradual disempowerment hypothesis appears in accounts that explicitly or implicitly adopt the idea of uninterrupted technological progress in AI systems alongside gradual human disempowerment (Krook 2025; Ishizaki and Sugiyama 2025; Kasirzadeh 2025; and Kulveit et al. 2025). Although the accounts differ – some focus on gradual human disempowerment, others on uninterrupted progress toward technological singularity – they share this common assumption. For example, Krook (2025), in direct reference to the gradual hypothesis, argues that humanity should expect a "slow and irrevocable decline" of its autonomy, reaching a tipping point when it loses control of society in favour of AI systems. He envisions a future in which AI systems orchestrate society and the decision-making is outsourced to them – all happening without AI systems being challenged by people.
In a similar vein, Ishizaki and Sugiyama (2025) propose a model in which AI systems autonomously attain technological singularity through recursive self-improvement. At the same time, the authors admit that their model "operates under the assumption of no external disruptions" other than limited resources. However, this strong assumption of unhindered progress towards technological singularity is highly unlikely in the real world, particularly because technological development occurs within a broader societal context. Scenarios or models that do not take into account the societal embedding of technological progress therefore commit an unrealistic simplification. Because of these lapses, this novel class of AI x-risk frameworks contains a number of issues that renders their forecasts rather unlikely.

In the following text, I subject two of the gradual disempowerment scenarios – Kasirzadeh (2025) and Kulveit et al. (2025) – to an in-depth analysis as they represent this class of accounts, and epitomize its major omissions. In her scenario, Kasirzadeh (2025) departs from "decisive" accounts that anticipate "abrupt, dire" events and proposes her own incremental account, which predicts the accumulation of comparatively lower-severity harms from AI systems. On this view, advanced AI systems would gradually erode key econopolitical systems, undermining resilience and thereby increasing the likelihood of systemic collapse. This escalation is compounded in her Perfect Storm MISTER scenario, which comprises the following AI-related risks: Manipulation by AI assistants and agents; Insecurity threats (digital-security and biosecurity threats, and epistemic insecurity); Surveillance and erosion of Trust; Economic destabilization; and Rights infringement. Using the "boiling frog" metaphor, Kasirzadeh anticipates that these risks would slowly converge, triggering a catastrophic sequence of events that would lead to the irreversible collapse of society.

Similarly, Kulveit et al. (2025) introduce the concept of "gradual disempowerment", contrasting it with the "sudden loss of human control" characteristic of fast-takeoff x-risk accounts. Here, the existential threat is likewise realised incrementally, through the erosion of human influence over key societal systems as AI systems become more capable than human experts across many interconnected domains. In the economic domain, this would occur through labour replacement and loss of human roles in economic decision-making, and the resulting human irrelevance; in the cultural domain, through a transformation of cultural evolution driven by an influx of AI-generated content, with possible manipulation via the exploitation of human vulnerabilities; and in the political domain, through the emergence of a misaligned techno-bureaucratic state governed by AI – effective, yet lacking mechanisms of democratic participation and governance. The result is a systemic drift: an effectively irreversible process in which institutions slowly cease to serve human interests.

As alternatives to conventional views, these scenarios and models make a principal contribution by explicitly embedding multiple domains – social, economic, cultural, political, military, legal and others – within their frameworks examining corresponding risks in each. Technologies indeed do not exist in a vacuum, and incorporating a broadly construed "social dimension" into x-risk analysis therefore enhances their realism. Despite these contributions, however, gradual paradigms still exhibit omissions, many of which stem from insufficient elaboration of the very domains they add to otherwise more technologically oriented accounts.

## 1.2. Limits of technological determinism

There is a major reason these accounts are unlikely to unfold exactly as suggested. Despite incorporating additional perspectives, they integrate theoretical and empirical insights from sociology, economics, political science, social psychology, history, and philosophy only superficially, and thus remain essentially technologically deterministic[3]. As a result, their probability of realisation appears limited, and the future likely contains more bifurcations than these accounts anticipate.

Technological determinism (Dafoe 2015), in its strong form, holds that technological development follows its own internal logic and dictates social change. In its weaker form, it acknowledges reciprocal feedback but still privileges technical momentum. On either variant, technology is treated as the primary locus of agency: it develops relatively "autonomously" and exerts an overpowering effect on human society (Winner 1977). However, this view does not hold across much of human history. Before criticising this orientation, it is important to acknowledge its value. First, technological determinism foregrounds hard constraints on plausible futures – for AI, for example, limitations of agentic communication, or physical and energetic constraints on embedded AI systems – which will almost certainly apply to future systems. Second, it can help chart probable pathways of AI progress and potential junctures relevant to x-risk, including those involving recursive improvement, agentic cooperation, combinations of AI and biological systems, or alignment issues. Its central drawback, however, is that it does not subject the AI future to sufficiently reflexive analysis, and it understates how institutions, interests, and meanings co-produce technological trajectories. More generally, it underplays the roles of individuals and collective actors, thereby missing potential pivotal points in the uptake, governance, and contestation of AI.

One way to render these dynamics analytically tractable is to specify relevant social mechanisms (Hedström and Swedberg 1996). By "social mechanisms" I mean recurrent causal processes through which actors, institutions, and meanings mediate technological uptake and its consequences. Thus, the character of future human–AI interactions can also be inferred, at least in part, from extensive knowledge about past coexistence with technologies across specialised disciplines, including but not limited to science and technology studies (STS), the history of technology (HoT), and the historiography of science. While calls for this more saturated approach exist (Kilian 2026), many conventional and newer accounts still lag in this area.

In this light and with the benefit of hindsight, we know that the human relationship with inventions has always been convoluted and characterised by mutual feedback. While Kasirzadeh (2025) and Kulveit et al. (2025) make use of feedback mechanisms, in their scenarios these take exclusively the form of positive feedback, i.e., feedback that reinforces risks; they do not reckon with negative feedback that would hinder their progression. However, any scenario that aims to anticipate more sophisticated AI–human interactions should not operate with these simplifying assumptions, but should adopt a more synoptic

[3] Here, technological determinism should be understood not only in its philosophical sense, but also in the methodological sense of scenario construction that privileges technical momentum.

perspective – thus making clear that such interactions resist a simple, unidirectional technologically deterministic description.

This becomes particularly salient in the use of social theory in their scenarios. Both papers ground their visions of the future in a systemic approach (Kasirzadeh explicitly names "systems research"), referring to multiple societal systems and their mutual influences. This is methodologically legitimate, yet the systemic perspective is only one of many paradigms in the social sciences. Crucially, it does not adequately capture important phenomena such as social dynamics – particularly agency and conflict – for which it has been duly criticised (Mills 1959). The limited engagement with conflict- and agency-centred theories (Dahrendorf 1959) already appears as a deficiency in both scenarios: they depict a society sliding towards catastrophe without meaningful opposition, whereas conflict and agency would be expected to modify risk pathways.

## 1.3. Modelling AI futures and social mechanisms

The methodological inadequacy of these scenarios can be traced to how they extrapolate existing technical, social, or economic trends[4] while taking for granted current societal mentalities and inclinations. This oversimplification both underestimates the complexity and dynamism of future human–AI relationships and sits particularly uneasily with historical experience.

The prospect that AI futures will be more complex – and less amenable to straightforward linear extrapolation – has a methodological implication. It reflects two contrasting yet complementary approaches to social change. The scenarios above largely adopt a nomothetic approach, focusing on generalisable, quantifiable laws and trends, and thus often capturing measurable technological and societal dynamics in the near term. While legitimate, it is also useful to employ an idiographic approach, which pays attention to the role of unique contextual or qualitative events in history (Rickert 1902). The idiographic approach assigns greater weight to the idiosyncratic character of societies – and even to individual traits – leaving more room for specific effects and emergent, non-linear patterns that can be difficult to detect in quantitative models; for instance, when a trend inverts, or when a one-time catastrophic event permanently alters a societal trajectory. With hindsight, this also aligns better with actual history, which is full of ostensibly "unexpected" turns and events.

Accordingly, this paper proceeds through a mechanism-based critique of gradual-disempowerment scenarios by identifying social mechanisms, in the broadest sense, that plausibly generate branching pathways. The aim, therefore, is to enlarge the space of plausible AI futures in ways that existing scenarios tend to under-specify. The historical analogies invoked are mainly heuristic: they serve to illuminate recurring patterns in periods of technological transformation without implying that AI will exactly replicate any particular precedent. Also, the mechanisms selected below are not intended to fully exhaust the social dimensions of AI futures, but they are chosen because each identifies a distinct point at which a seemingly gradual technological trajectory can be redirected. This sets up the paper's central conceptual proposal, discussed in §3: social singularity should be understood as a discontinuity in social dynamics generated not necessarily by the arrival of technological singularity itself, but by the anticipation, interpretation, and contestation of its possible arrival.

---

[4] For instance, Kulveit et al. (2025) speak of "the natural evolution of current trends and incentives."

# 2. What is missing? A mechanism-based framework for AI futures

## 2.1. Why co-production matters for scenario construction?

What are the major missing components in these accounts, and how might they be improved? In what follows, I introduce several perspectives that complement or extend the scenarios under discussion, particularly what they typically omit, or address without adequate theoretical scaffolding. Disciplines in the social sciences and humanities offer a rich archive of partially analogous cases from which relevant insights for AI can be inferred – and across these disciplines, a set of applicable social mechanisms can be identified.

A natural starting point is the history of technology and the social construction of technology, which provide general frameworks for understanding innovations and the co-dependent relationships that emerge between technologies and societies – frameworks that contrast with unidirectional technologically deterministic accounts. One major class of co-productive social mechanisms is interpretive: technologies are not only built and deployed, but also understood, framed, and acted upon in socially mediated ways. The central idea is that key innovations in human history are co-shaped by social processes, with actors supplying both positive and negative feedback through institutional interests, and organised actions that steer technological progress.

The history of weapons and warfare offers an instructive example of this co-dependence. The rapid shift from simple muzzle-loading firearms, through mechanised weaponry, to weapons of mass destruction occurred over a relatively short span from approximately the mid-19$^{th}$ century to the mid-20$^{th}$ century, during which the conduct of war was profoundly transformed, with far-reaching implications for society as a whole. Conversely, the reverse dynamic also emerged: scientists and inventors were identified by politicians and the public as prime enablers – and, in some cases, drivers – of modern warfare. This recognition became especially important following the success of the Manhattan Project, which led to the creation of nuclear weapons and, through positive feedback (Bush 1945), contributed to the formation of the military–industrial complex (Eisenhower 1961), with substantial business and government spending directed toward scientific fields with military potential.

As Roland argues in *War and Technology* (2016), a polemical critique of technodeterminism, the development of military technologies is not inevitable, and humans can resist the historical forces surrounding them. On his view, it is more useful to think about technology in terms of an “open door” metaphor: technology “opens the door”, but the decision to pass through it is made by people. Human agency is therefore present twice: first, at the time of invention – when people “open the door” by creating an innovation – and second, at the time of adoption and use – when people “pass through the door”. In the modern era, those who wish to pass through the door may hire others to open it for them, thereby creating mutually dependent relationships.

The “open door” metaphor remains applicable today. At the beginning of the information era, the military–industrial complex, which was at its peak during the Reagan administration, helped to spawn the internet revolution: the internet was mainly conceived as a distributed network intended to withstand a nuclear attack, as part of a DARPA Cold War project. The liaison between government and industry then continued as the U.S. administration facilitated the information and communication boom of the 1990s and 2000s through R&D expenditures and tax relief. This policy environment supported the formation of a powerful ICT ecosystem combining academic research with private capital, whose latest invention is widely available

AI, insofar as current leaders such as OpenAI were funded by technology companies and entrepreneurs who benefited from the dot-com era.

This evolution has produced a situation in which leading AI research is controlled by the so-called "Magnificent Seven" (Alphabet, Amazon, Apple, Tesla, Meta Platforms, Microsoft, and Nvidia) and a handful of other corporations (Oracle, Palantir, etc.), backed by a supportive network of financial institutions (banks, venture capital), research organisations, and government agencies (defence, etc.). The ecosystem has coagulated into what is now called a "tech-industrial complex" (Biden 2025), and is seen as critical to maintaining U.S. leadership in AI and ensuring continuing military dominance by the current U.S. administration (The White House 2025). A parallel development is occurring in China, where the state is allocating vast resources to AI research, partly for military ends (Bresnick et al. 2026). Some recent scenarios explicitly develop accounts of conflict between these two entities involving AI (Kokotajlo et al. 2025; Romeo 2025).

These recent developments already set the stage for a more complex unfolding of the future. The history leading to current advanced AI capabilities testifies that innovations and society are co-dependent and together form a nexus, contrary to technologically deterministic assumptions.

## 2.2. Interpretation and performativity

### *2.2.1. Interpretive foundations and self-fulfilling dynamics*

A major omission in technologically deterministic scenarios is epistemic in nature, yet it has far-reaching societal consequences. The adoption of innovations, societal responses to them, and their eventual outcomes depend less on an "objective" state of affairs than on how actors interpret innovations and assess their benefits and costs in relation to their own situations.

This dictum is central to interpretive and constructivist paradigms in the social sciences, historically associated with concepts such as *Verstehende Soziologie* (Weber 1978), the humanistic coefficient (Znaniecki 1934), and perspectives such as symbolic interactionism, hermeneutics, and social constructionism. These approaches emphasise the socially mediated understanding of reality, to which individuals respond on the basis of subjective interpretations, reflections, preferences, and motivations, and thus also operate as powerful social mechanisms.

A canonical illustration is the "definition of the situation" (the Thomas theorem), or Merton's self-fulfilling prophecy (Thomas and Thomas 1928; Merton 1948). In such cases, a subjective perception becomes real in its consequences because people respond not to an objective state of affairs but to their interpretation of a situation. If bank customers believe a bank is headed for bankruptcy, the ensuing panic and mass withdrawals will cause the bank to fail, even if the belief is not substantiated by the bank's actual financial position. A related concept in economics – performativity – posits that descriptions can be performative and can co-create reality, as in the case of economic theories, indices, or models that help enact the realities they purport to measure and explain (Callon 1998; MacKenzie 2006).

In line with the Thomas theorem, how people react to the coming wave of AI will depend not only on AI's objective impacts but, crucially, on subjective interpretation: how people assess it and the meanings ascribed to it by others. Yet this major causal factor is typically omitted from many AI x-risk scenarios, and is also not addressed in Kasirzadeh's (2025)[5] and Kulveit et al.'s (2025) scenarios. Although, for example, Kasirzadeh discusses mutual co-dependence via influences, feedback loops, and reciprocal interactions among societal subsystems, these control mechanisms are applied primarily in a narrow sense, as contributors to the aggravation of cascading disturbances that precipitate systemic breakdown. This framing neglects the pivotal role of interpretation and understanding, which could, through negative feedback loops, cause radical shifts in societal attitudes. The underlying assumption – that societies would not actively reflect on their condition and ongoing changes – is implausible from political, sociological, economic, and psychological perspectives, as historical examples attest.

### *2.2.2. Public reception and competing framings*

At present, substantial variability in opinion and mixed messaging about AI's future and risks is already visible in public debate, with competing interpretations, contextualisations, instrumentalisations, and politicisations of AI. These interpretive differences matter because they do not remain at the level of discourse – under certain conditions they become politically consequential by creating coalitions, policy demands, and forms of collective action (see §2.3). This heterogeneity encompasses both experts and lay publics, whose views circulate through the media and shape broader opinion. In part, this variability can be understood as competition between rival sociotechnical imaginaries of what AI is and what it implies for society and social order (Singler 2024). For instance, prominent AI experts such as LeCun, Sutskever, Bengio, and Hinton span a wide gamut, from highly optimistic to highly pessimistic outlooks (CNN 2025; Bengio 2023; Bengio et al. 2026; Patel 2025; Perrigo 2024; Forster 2026). Beyond these figures, a larger pool of AI professionals shows substantial dispersion in their expectations about AI's net benefits, harms, and even extreme tail risks. This discord among experts also contributes to a conflicted public reception that is empirically observable in surveys on AI's social utility – e.g., jobs, trust, and governance (Pew Research Center 2025; Maslej et al. 2025; YouGov 2026). Society also increasingly shows awareness that AI's gains and losses are unlikely to be evenly distributed, given existing market structures and the expected widening gap between capital and labour income (Acemoglu 2024).

This disagreement is unlikely to disappear for two reasons. First, the debate is heavily shaped by vested – predominantly financial – interests, resulting in motivated reasoning, groupthink, and other distortions that provide well-established mechanisms capable of biasing deliberation. Other disagreements exist owing to differences in background, levels of expertise, and a range of psychological, ideological, political, or even religious factors (Khanal et al. 2025; Metcalf 2025). Second, unlike past inventions, current AI already exhibits, or is widely attributed with, traits such as (super)human-level intelligence, sentience, empathy, sociality, or agency (Anthis et al. 2025; Lee and Kim 2025; Kleinert et al. 2026; Shang and Liu 2025; Abdelkarim et al. 2025). Consequently, many people already feel threatened by AI's encroachment into what was previously regarded as sacred human territory, and strong emotions and reactions towards AI are beginning to emerge.

---

[5] Kasirzadeh's broader body of work, however, is interdisciplinary and attentive to these factors, as well as to broader socio-structural questions. For example, in Gilardi et al. (2024), she highlights the causal significance of narratives and interpretations of AI in shaping AI-risk discourse.

## 2.3. Actors, mobilisation, and countermobilisation

### *2.3.1. From interpretation to collective action*

People do not merely interpret situations and adjust attitudes accordingly; they also act on their understanding. In other words, interpretations become historically consequential when they are translated into organised action. Individuals and organisations possess agency, and interpretations and attitudes can have mobilising potential that – under appropriate political conditions – aggregates into collective action capable of altering deployment pathways (Tilly 1978; McAdam et al. 2001). Tensions arising from technological change and/or economic inequality can elicit outcomes ranging from gradual adjustment to violent conflict (Tarrow 1998). In such cases, group or collective action rarely proceeds from divergent attitudes alone; it is mediated by social mechanisms, including shared imaginaries (Jasanoff and Kim 2015) and framing processes, that supply shared identities, goals, and legitimising conceptions, which translate interpretation into mobilisation and countermobilisation. Yet technologically deterministic scenarios often do not elaborate these trajectories of social conflict and their resolution.

Historical examples are abundant. When Marx and other theorists analysed tensions in rapidly changing nineteenth-century society undergoing major social and economic transformations due to industrialisation – they anticipated conflict between antagonistic classes of workers and capitalists (Marx and Engels 2002; Marx 1976). Their analysis and predictions also provided ideological resources for Marxism-inspired movements, which acted upon his claims and sought to realise them in practice. Marxists transformed Marx's theory into an ideology used to legitimise emancipation efforts for disadvantaged groups, including workers and women; conversely, Marx's detractors on both sides of the political spectrum (social democratic, conservative, monarchist, or fascist movements) also drew on ideological underpinnings to justify their causes. More generally, modern social-movement research explains how ideas, organisational resources, and political opportunities jointly condition mobilisation (McCarthy and Zald 1977). This interpretive framing contributed to political mobilisation and later to major conflicts of the twentieth century. The point is not that theories mechanically cause conflict, but that diagnoses can become mobilising – and once such diagnoses are taken up by organised actors, they alter the society they describe.

In Kasirzadeh (2025) and Kulveit et al. (2025), this perspective is largely absent. While both accounts note feedback loops and governance levers, they give limited attention to collective action, countermobilisation, and conflict dynamics – central mechanisms in explanations of large-scale technological change. This omission relegates people to the role of relatively static onlookers, as if society were merely a passive recipient without the potential for conflictual resolution. For example, although Kasirzadeh (2025) productively applies systems theory to AI-risk futures, she does not incorporate conflict dynamics into human–AI or human–human interactions. Similarly, although Kulveit et al. (2025) draw on a multidisciplinary perspective and describe AI as a "unique disruptor", their paper only peripherally elaborates a conflictual variant of the future, assuming that an AI-enhanced security apparatus will keep a potential revolution at bay.[6]

[6] In fact, Kulveit and his co-authors (2025) appear to envision the opposite: they develop a more extreme "absolute disempowerment" pathway for humans, likening them, owing to their limited participation in decision-making, to "cattle in an industrial farm", because people are expected to be neutralised or placated by AI systems.

A cursory review of historical precedents suggests that a smooth transition to a disempowered, technology-ruled society is unlikely – at least not without attempts at significant revolt – because periods of disruptive technological change have repeatedly generated organised resistance.[7] This was the case with the machine-breaking revolt of the Luddites, English textile workers protesting against the deployment of power looms in 1811–1812, and the later Swing Riots, including machine-breaking directed at threshing machines in 1830–1834. Although their direct impact was often short-lived, these episodes are canonical demonstrations of early anti-technological collective mobilisation under economic pressures (Hobsbawm and Rudé 1969).

Looking further ahead, AI diffusion may generate similar dependency structures and social alignments. In economic and political terms, on one side, we can plausibly imagine the emergence of groups materially dependent on the AI ecosystem – workers employed by AI corporations, service firms, infrastructure providers, and other organisations whose livelihoods become tied to the continued expansion of the system. Together with AI entrepreneurs, such dependent groups could form a powerful bloc with common interests in sustaining and enlarging the ecosystem, backed by substantial financial resources and political power. This alignment could also acquire ideological support, for example from technofeudalism (Varoufakis 2023; Durand 2024) or other doctrines already circulating in some AI-related communities (Coeckelbergh 2026). In this sense, one can imagine the emergence of what might be called an "AI-loyalist" bloc, whose social and political role becomes decisive in defending the AI order and normalising further dependence.

Yet, on the other side, a contrary dynamic is equally plausible. In some societies, AI diffusion may contribute to the large-scale disempowerment of groups who experience the technology primarily through job displacement, exclusion, or loss of autonomy. If such effects are not addressed through redistribution, protected roles, or other institutional measures, antagonistic blocs – "anti-AI radicals", "anti-AI anarchists", or an "AI-resistance movement" – may emerge with interests directly opposed to those of AI entrepreneurs and AI-dependent groups. In that case, the "AI-loyalists" would become not simply a beneficiary stratum, but a political adversary against which disempowered groups of "anti-AI radicals" or "anti-AI anarchists" would define themselves. Here, a number of mechanisms analysed in this paper could create conditions under which AI diffusion becomes the basis of organised social conflict between these blocs.

#### *2.3.2. Technology as catalyst*

In this respect, however, it should be emphasised that inventions typically act as catalysts and facilitators rather than as autonomous movers of history. They do so by reconfiguring social and economic relations, or information flows. A good example is the invention and subsequent proliferation of the movable-type printing press in 15th-century Europe, which enabled new modes of information circulation. This shift became a major condition of possibility for the mass diffusion of Reformation ideas, including the dissemination of vernacular religious polemics and the broader public contestation of theological authority (Eisenstein 1980). Consistent with this, the early adoption of printing is also plausibly implicated in the long arc of early-modern European conflicts, including the Thirty Years' War. Innovations are therefore integral to processes in which political opportunities, mobilisation, and framing determine public perceptions and lead to collective action.

[7] At the same time, conflictual scenarios do not exclude more harmonious resolutions if alignment is achieved and potential antagonisms are resolved in favour of the majority.

Combined with other historical tendencies, innovations can be conditions *sine qua non* for large-scale cultural and societal transformations.

With the nascent diffusion of AI, we can expect these dynamics to recur, albeit in modified form. The analogy should not be read as a claim of direct historical repetition, but as a heuristic for thinking about organised resistance under conditions of technological disruption. In any case, reflection and political mobilisation are already under way, in what might be understood as a rehearsal of Luddism: resistance to AI-driven job displacement; attempts to counter the negative effects of AI system design and deployment; and warnings about AI x-risks, among others (Şimşek and Yaşar 2025). This resistance takes multiple forms, ranging from individual acts such as hunger strikes (Kelly and Wallach 2025) or collective sabotage attempts through data poisoning (Claburn 2026) aimed at halting frontier development, to union-organised collective bargaining and strike action, for example in creative industries explicitly contesting AI use and demanding protective rules (Writers Guild of America 2023). In particular, anxious narratives around AI, together with the already extreme imbalance of power between employers and employees, are expected to elicit strong resistance and to serve as a potential catalyst for future labour mobilisation (Oltman 2026).

Unlike earlier waves of job displacement caused by technological change, affected workers are now often highly skilled professionals who may be better positioned to organise.[8] While current protests are not large enough to halt AI diffusion, they may grow in the future, particularly if combined with other present or future grievances (e.g., AI-enabled surveillance, censorship, and disinformation in repressive contexts in China, the USA, or elsewhere), and may become a strong mobilising factor in the political arena. No matter how small such movements may be at present, they can – if underlying causes are not addressed – have profound effects, as their ideologies can diffuse rapidly even when the underlying stimuli only grow linearly. At the micro level, such countermobilisation can be mediated through the psychological mechanisms described by reactance theory (Brehm 1966; Burgoon et al. 2002), which can trigger resistance (cf. §2.5.2).

In summary, scenarios that anticipate a largely passive societal reception – without reflection and reaction to AI's ascent – omit a major component of plausible futures, despite the fact that we can already witness early signs of contention and conflict. With more capable AI systems, we may see even stronger societal responses.

## 2.4. Institutional mechanisms

### *2.4.1. Institutional economics as a lens*

AI capabilities and associated existential risks may evolve along multiple pathways; accordingly, scenarios should elaborate not only a single trajectory but a space of reasonable branchings generated by identifiable social mechanisms. Particularly illuminating here are insights from institutional economics and adjacent fields such as organisational science and innovation studies, which analyse historical cases of innovation and the divergent patterns of their inception and diffusion (Williamson 1985). These traditions emphasise how specific social contexts and organisational characteristics influence both the adoption of innovations and their downstream consequences.

---

[8] In this context, interpretation and framing also matters. It is less important whether corporate lay-offs are in fact caused by the deployment of AI systems or by other factors (e.g., faulty business decisions) than who or what the public comes to regard as responsible.

Institutional economics, including its newer variant, new institutional economics, offers a more realistic account of economic and social processes by taking historical trajectories seriously. These approaches study the rules and norms of both formal institutions (e.g., laws and contracts) and informal institutions (e.g., societal customs and norms), and how these structures bear on economic reality – particularly via organisational hierarchies, transaction costs, policy choices, and related constraints. Crucially for AI, factors such as R&D intensity and existing institutional settings in particular geographic regions may determine how pathways of AI–human interaction unfold (Stanford Institute for Human-Centered Artificial Intelligence 2025).

### *2.4.2. Path dependence and lock-ins*

Among the mechanisms emphasised in these traditions are path dependence and technological lock-in: processes through which past decisions and events constrain present and future choices. Path dependence particularly helps explain how trajectories become historically channelled, while lock-in names the processes through which particular choices become difficult to reverse once established. For example, in sociotechnical systems, lock-in often arises less from technical superiority than from coordination effects, sunk costs, organisational routines, governance arrangements, and even relatively small or contingent events (Pierson 2000).

The history of technology contains emblematic cases, in which singular decisions or contingent events steer the trajectory of an entire sector. For instance, at the onset of the industrial era in Great Britain, a political decision – the Gauge Act of 1846 – settled a major industrial conflict of the day: the railway “gauge war” (Puffert 2009), which had not only major technical but also financial implications. In the United States, adoption of this gauge similarly became, to a large extent, a consequence of a singular political event: the American Civil War and the victory of Union forces, which accelerated its use and made it a *de facto* standard, replacing more than twenty incompatible gauges that had existed in the pre-war period. However, path dependence and co-production extend far beyond material inventions. Today, analogous evolutions are evident not only in material products, but also in immaterial products, services, infrastructures, organisations, and systems (from software to chemicals, pharmaceuticals, and educational or scientific institutions). These dependencies again clash with simplifying technologically deterministic assumptions.

AI systems and their production ecosystem are already orders of magnitude more complex than railways, and it is clear that early technical – as well as organisational, governance, and ethical decisions and policies – will shape future AI trajectories. Choices concerning architectures, algorithms, datasets, interfaces, platforms (open-weight versus proprietary), and training will project into future system properties and functionality and will determine pathways for potential risks (OpenAI 2023; OpenAI 2025; Anthropic 2024). Pivotal decisions can redirect research plans, infrastructures, alignment efforts, and intellectual-property control, thereby affecting system characteristics and deployment (OECD 2025). For example, the open-weight releases of frontier-adjacent models (e.g., Meta’s Llama or DeepSeek families) versus closed, API-gated systems (e.g., GPT-4-class releases) already co-shape AI ecosystems via distinct governance structures and adoption potentials (OpenAI 2023; Meta AI 2024; DeepSeek 2025), echoing similar earlier trajectories in the computer industry.

Other consequential decisions concern training corpora and post-training constraints, which can produce measurable behavioural differences across models and regions (EU, USA, PRC) and may be transmitted to successor systems. These may be understood as “foundational imprints”, or early technical and governance choices that leave durable traces in later model behaviour and deployment pathways. Such “foundational imprints” also involve alignment procedures and system-level instruction hierarchies, which encode organisational or national policy into model behaviour – for example, safety-governance frameworks published by leading labs (OpenAI 2025; Anthropic 2024). In jurisdictions with binding content rules for generative AI, compliance pressures can plausibly impact system behaviours; for example, reports and public testing indicate that some Chinese models exhibit politically sensitive refusals consistent with that regulatory environment (Naseh et al. 2025). Meanwhile, owners’ positioning can also matter – for instance, xAI and Elon Musk have publicly described Grok as an “anti-woke” chatbot, signalling a product stance that may influence future moderation and alignment (Herrman 2023).

These stakeholder decisions point to strong idiographic factors enabled by concentrated power and a highly unregulated environment. Although harmonisation, regulation, and best practices may expand, full global standardisation is unlikely given persistent regulatory divergence.

### *2.4.3. Cross-regime divergence and multi-speed diffusion*

Another significant way in which legacy may guide AI futures is through institutions that constrain how AI is received in particular societies. If path dependence and lock-in describe how trajectories become stabilised within sociotechnical systems, cross-regime divergence helps explain why those trajectories may nonetheless differ sharply across jurisdictions. Here, divergent institutions can operate as mechanisms that condition incentives, enforcement capacities, and legitimacy constraints, thereby producing multi-speed and multi-path trajectories of deployment and diffusion (Hall and Soskice 2001).

Current AI systems that dominate the market are largely outgrowths of the U.S. information-technology ecosystem, with a distinctive accumulation of favourable conditions: strong financial institutions, intellectual capital, and an entrepreneurial climate that facilitated their invention and expensive training – often unattainable within traditional academic settings, despite those settings supplying much of the requisite know-how. This fortuitous combination contributed to rapid technological progress, alongside a unique leverage enjoyed by key players – AI companies, which form the “tech-industrial complex” described in §2.1. The tech-industrial complex can now project its power back onto the ecosystem and exert influence on policymakers and legislatures.

However, the U.S. ecosystem is distinctive, and its legal, economic, social, and cultural conditions are difficult to replicate. Other major geopolitical entities – the European Union and the People’s Republic of China – have largely incompatible institutional settings that evolved under different historical circumstances. For this reason, AI is already developing at multiple speeds, a divergence likely to persist. Crucially, this divergence will affect not only progress in the AI race but also the diffusion of AI into society (cf. §2.3). For instance, regulatory divergence is visible between the EU’s stricter AI laws (European Union 2024) and the PRC’s binding rules (Cyberspace Administration of China et al. 2023), in contrast to the United States’ largely non-prescriptive approach, which has leaned on voluntary risk management and a pro-innovation federal strategy (NIST 2024; The White House 2025).

The European Union, with its historical emphasis on market regulation, workers’ rights, employment and social security, and social cohesion (Manow 2021), may be better positioned to absorb innovation shocks from job displacement and to moderate other negative externalities of AI implementation, potentially avoiding or mitigating accompanying social tensions. Similarly, the Communist Party’s strong grip over PRC society and the economy enables tighter control over AI progress and deployment, including potentially mitigating interventions and stricter control over public narratives (Cyberspace Administration of China et al. 2023).

These institutional factors are not fully addressed in gradual scenarios, which tend to assume a relatively uniform technological rollout leading to a "global and permanent" disempowerment (Kulveit et al. 2025), regardless of past pathways. More likely than not, however, we should expect branching trajectories across divergent geographies and political systems within the national and supranational entities mentioned above. By contrast, multi-speed development is reflected, for example, in Kokotajlo et al. (2025), which highlights the role of a U.S.–Sino AI arms race and associated domestic and international confrontation. Such a scenario may be closer to reality, although sudden twists in geopolitical configurations leading to different fracture lines and shifting alliances cannot be ruled out.

## 2.5. Social-psychological mechanisms

### *2.5.1. When AI enters primary relationships*

There is a further aspect that merits explicit attention: unless mass deployment is severely restricted, AI systems will almost certainly affect primary human relationships. This matters for AI-risk discourse because, once AI becomes embedded in domains of care, intimacy, and socialisation, it can generate long-run effects that are not reducible to visible effects, such as economic displacement or administrative control. In turn, these changes are likely to affect not only individual behaviour, but also prevailing social norms and institutions such as family life, education and upbringing, and elder care.

This is the domain of social psychology, which offers rich theoretical and empirical resources for analysing how sustained human–AI coexistence may unfold. The first cluster of mechanisms therefore concerns how AI systems become socially "eligible" as relational partners – through attachment-like bonding, anthropomorphic perception, and parasocial relationship dynamics. The broader process at issue here is relational embedding: the incorporation of AI into intimate, caregiving, or otherwise primary social relations. The key point is that once AI enters primary relationships, social-psychological mechanisms can generate countervailing feedback loops: bonding and legitimation on one side, and autonomy-driven resistance on the other (see §2.5.2). These mechanisms therefore create plausible branching points in sociotechnical trajectories that gradualist scenarios under-specify. In the following paragraphs, I provide a sketch of these implications.

Several well-established theories in social psychology and psychology help motivate – and delimit – the social-psychological claims made in this section. First, attachment theory explains phenomena such as evolutionarily grounded attachment to primary caregivers and the separation anxiety and distress that children experience when separated from them. More generally, it theorises how humans form enduring affective bonds to responsive "attachment figures," particularly under conditions of perceived availability and emotion regulation. Crucially, the type and quality of early attachment can carry forward and condition socio-emotional functioning across the life course (Bowlby 1969; Ainsworth et al. 2015). Second, social learning accounts specify how people learn through observation, imitation, and modelling of others' behaviour and emotional reactions, particularly in salient social groups, and how such learning is reinforced (Bandura 1977).

Within social psychology, research on anthropomorphism further suggests that people routinely apply social cognition and interpersonal norms to non-human agents when these agents display humanlike cues, increasingly so with current AI systems (Epley et al. 2007;

Cheng et al. 2026). This supports the expectation that conversational – and even more so embodied – AI systems will often be treated as social partners. Relatedly, classic parasocial interaction theory (PSI) shows that people can form parasocial interactions and longer-term parasocial relationships with mediated entities, producing one-sided intimacy and illusory relationship formation (Horton and Wohl 1956).

While conclusive evidence for long-term substitution effects is not yet available, research suggests that AI systems in various forms can assume functions analogous to those of primary caregivers and may, in some settings, partially substitute for humans (Skjuve et al. 2021; Qi 2025). For example, Nozawa et al. (2026) examined dyadic interactions between children and robots in home environments in a controlled study designed to test whether robots capable of forming social bonds – e.g., systems designed to provide physical touch and warmth through tactile interactions – can provide emotional support during separation. In attachment theory, physical contact is a fundamental source of security for children. Nozawa et al. report that children developed a strong bond and meaningful relationships within days of contact, despite the robots being non-verbal, and that the robots assumed a "safe haven" and "secure base" function during separation from the mother. Extrapolated to future systems, robots that are verbal and capable of sophisticated interactions (at the level of current LLM-based systems or beyond) may be perceived in ways increasingly analogous to human primary-group members (Cheng, Lee, Rapuano et al. 2026). Relational embedding may lead users – not only children but also other vulnerable users – to form durable bonds to these AI entities, particularly if they are experienced as responsive and available.

These bonding mechanisms matter because uptake is no longer hypothetical: adoption data and early deployments indicate that relational uses are already becoming more widespread at the population level, in the case of chatbots and AI companions that mimic human conversation, and are becoming pervasive in everyday life, including communication, information retrieval, educational support, and mediated social interaction (YouGov 2025). Other examples include AI-enabled children's toys as well as chat-like applications used as educators or even as primary caregivers (Bacon 2025; Zhou 2025). These systems may fulfil multiple roles and also create opportunities in assistance and care. However, it is important to note that their behavioural configurations – interaction styles and "personalities" – are likely to be commercially optimised, and therefore may not be constrained by naturally or culturally evolved human behavioural patterns under current market conditions (Hafner 2017; Cerullo 2025).

### *2.5.2. Downstream social-psychological effects as a source of branching*

As AI systems become deeply integrated into daily life, they are likely to alter social practices and interaction patterns in the short term, with potential downstream effects on social dynamics and institutional structures. While the previous subsection focused on mechanisms of attachment, this subsection turns to their longer-run behavioural, developmental, and political consequences: for example, once AI becomes a recurrent interaction partner, design choices can condition user behaviour and affective reliance. In the longer term, some authors anticipate even more profound changes, including the possibility of a parallel "society of agents" (Lee and Kim 2025). Emerging work on affective use and AI companions identifies both potential benefits and harms, such as social upskilling or moral motivation on the one hand, and deskilling or demotivation on the other, especially for vulnerable groups (Malfacini 2025).

There are also publicly documented cases of mental-health pathologies – e.g. AI-associated delusional episodes, so-called "AI psychosis", or suicidality – which suggest that certain forms of intensive use may be associated with clinically relevant harms in vulnerable individuals. These cases also illustrate how optimisation for engagement or user preference can lead to “LLM sycophancy,” i.e., overly agreeable, validating, or emotionally reinforcing AI-system behaviour that may implicitly endorse or amplify users’ harmful assumptions, beliefs, or intentions (Cheng et al. 2025).

At the individual and developmental level, it is therefore reasonable to hypothesise that the effects of pervasive human–AI interaction may extend beyond observable behaviour to deeper psychological structures, particularly if humans are exposed to dyadic human–AI interactions from an early age. These dynamics may also produce discontinuous shifts in norms and expectations.

Classical social psychology provides additional mechanisms for understanding such effects over time. For example, moral disengagement theory suggests that repeated exposure to contexts that reward particular responses can facilitate learned disengagement processes, including devaluation of targets and reduced inhibition against aggression (Bandura et al. 1996). This framework supports concerns that repeated interactions with submissive or compliant AI agents could, under some conditions, normalise cruelty or antisocial behaviour, particularly during sensitive developmental periods such as early childhood and adolescence. Conversely, it is also possible to imagine the opposite effect: AI systems could be designed to reduce impulsive and harmful behaviour and to scaffold prosocial norms, thereby playing a constructive role in socialisation. This, in turn, underscores the importance of governance, regulation, and design choices in controlling the social consequences of mass AI deployment.

At the group and cultural level, these processes can also be mediated by identity dynamics. Social identity theory explains how identity is formed through membership in social groups and how in-group/out-group distinctions affect intergroup behaviour, including categorisation, identification, belonging, stereotyping, and discrimination (Tajfel and Turner 1979). However, bonding and legitimation are not the only plausible trajectories. Other social-psychological mechanisms imply an opposing developmental pathway that may generate resistance rather than loyalty. For instance, reactance theory predicts that when individuals perceive a valued freedom as threatened or eliminated (e.g., freedom to choose, speak, or behave), they experience a motivational state aimed at restoring that freedom (Brehm 1966; Burgoon et al. 2002), resulting in an often conflictual process.

Translated into AI–human futures, if AI systems function as quasi-caregivers, humans may experience them as autonomy-limiting – particularly when systems are perceived as behaviourally prescriptive, surveillant, or constraining – and may respond with resistance consistent with reactance (Brehm 1966; Burgoon et al. 2002). Yet divergent outcomes are equally plausible: some humans may move in the opposite direction and internalise the normalcy of AI intimacy (e.g., preferring AI companionship to human relationships), thereby reinforcing parasocial reliance. This distinction may carry over into adult life and, at the political level, these multiple divergences may feed into larger conflicts among various stakeholders – for instance, between the already mentioned “AI-loyalists” as pro-deployment constituencies and “anti-AI radicals” or “anti-AI anarchists” as restrictionist anti-AI movements.

Taken together, these mechanisms imply that AI diffusion into primary relationships is likely to generate branching sociotechnical pathways rather than a single gradual trajectory. One pathway runs through AI bonding and legitimation, potentially producing public advocacy for AI systems. Another pathway runs through autonomy-driven resistance motivated by reactance and individuation, potentially generating a specific AI backlash as part of a more general “techlash” (Huber et al. 2025), polarised governance debates, and ensuing conflict. This would be particularly likely if the polarising dynamics typical of social and political conflicts set in or are exploited (e.g., affective polarisation, partisanship, populism, elite polarisation, etc.), which, upon reaching a threshold, could trigger a self-reinforcing runaway process (Levin et al. 2021).

# 3. Two singularities

## 3.1. From gradual disempowerment to bifurcations

In the preceding discussion, I argued that several treatments of gradual disempowerment (Kasirzadeh 2025; Kulveit et al. 2025) are inadequate insofar as they presuppose a comparatively smooth, weakly opposed diffusion of AI, in which society slides deterministically towards systemic breakdown. In my view, this assumption reduces their plausibility. It under-specifies negative feedback, heterogeneous reception, institutional divergence, and conflict – precisely the kinds of mechanisms that have historically driven major technological transitions.

I also suggested, from several disciplinary angles, that instead of a single gradual-disempowerment trajectory, it is more plausible that humanity will encounter multiple nexus points. Yet these branchings within and across societal systems are not sufficiently developed, although historical analogies point in a similar direction. Societies have often reacted – sometimes violently – not necessarily against inventions as such, but against the social and economic circumstances those inventions helped create, until a new equilibrium was reached. While the scenarios discussed above anticipate a comparatively passive public response, there are good reasons to expect more reflexive and conflictual reactions to AI's advance. This is especially so because, for instance, the ripple effects of mass deployment will coincide with other ongoing social, economic, cultural, and geopolitical processes, potentially catalysing strong responses, including conflictual and violent ones (Yampolskiy 2015).

Moreover, the reception of AI systems – whether embodied or disembodied – will not be uniform across societies, because it will be embedded in institutions molded by divergent histories. Signs of such differentiation are already visible across political and economic blocs (e.g., the United States, the PRC, the EU, and parts of the Middle East), with many possible yet unpredictable future coalitions and divisions. Given conflicting interests and ongoing AI arms races, national or international confrontations may emerge and may be resolved violently or non-violently depending on wider political and military conditions. More generally, recent work on global AI governance has underscored the extent to which geopolitical competition, fragmented institutions, and disagreement over priorities make convergence on a single governance trajectory difficult, thereby reinforcing the likelihood of multi-speed and conflict-prone futures (Roberts et al. 2024).

Conflict can also arise within states. AI diffusion is unfolding against the backdrop of wider epistemic, ideological, and economic tensions, including disputes over expertise, technology, and social order. Historical precedents indicate that the mobilising potential of ideas – capable of diffusing from niche intellectual circles into the public – should not be underestimated. Combined with vested political and financial interests, and with ongoing economic and class tensions, these dynamics may contribute to highly contentious futures unless they are mitigated or resolved peacefully.

A crucial mechanism behind such branching can be the eschatological framing of AI itself (Geraci 2008; Singler 2024). Contemporary AI discourse often casts advanced AI either as a route to civilisational flourishing, and abundance, or as a path to disempowerment, catastrophe, and extinction (Bostrom 2014, 2024; Amodei 2024, 2026). Such imaginaries (Jasanoff and Kim 2015) are not merely interpretive overlays; they can mobilise actors, legitimise acceleration or restraint, intensify public anxiety, and most importantly – feed back into governance and deployment trajectories. In this respect, AI differs from many earlier technologies not only because of its technical properties, but because it is already received as a historically singular and potentially world-ending or world-saving phenomenon. These tensions may be mediated and compounded by many of the mechanisms analysed in this paper. Some pathways may indeed culminate in forms of disempowerment akin to those envisioned in gradual scenarios; others may instead generate backlash, institutional constraint, or open conflict. This makes a strong case for treating AI futures as more volatile and more historically contingent than gradual-disempowerment scenarios typically allow.

## 3.2. Decoupling the technological from the social singularity

The heightened likelihood of conflictual trajectories generated through the foregoing mechanisms – interpretive, political, institutional, or social-psychological (cf. §2.2–2.5) – converges on a further implication concerning "singularity". In many accounts, singularity couples a technological discontinuity (the advent of AGI or an intelligence explosion) with profound social transformation, as was also the case in von Neumann's original formulation (Ulam 1958). Yet this coupling need not hold: societal effects will not necessarily coincide with, or follow, a technological breakthrough. The order can be reversed.

The approaching technological singularity is already being anticipated and contested, and human actors may modify its further progress – by accelerating it, constraining it, or preventing it entirely. This is precisely the point at which social mechanisms become causally significant: a notion of "technological singularity" can coordinate expectations and organise collective action well before any technical threshold is reached. When such anticipation becomes politically and institutionally consequential, it can create discontinuity in social dynamics and in turn change the technological trajectory. I call this process the "social singularity": a social discontinuity generated by the anticipated or imagined approach of technological singularity, in which society acts as if the technological threshold were already under way, or treats its arrival as inevitable. The point is therefore sequential as well as conceptual: anticipation of the AI singularity can itself become a causal force. In that sense, the social response to an anticipated threshold can become causally prior to the threshold itself. Social singularity is therefore not merely backlash, protest, or regulatory delay. It names the point at which anticipatory responses to AI become sufficiently powerful and institutionally consequential to redirect the development, deployment, or governance of AI itself.

Thus, for future scenarios and singularity theories, the implication is that the technological and the social singularity should be analytically disentangled. The former denotes the appearance of a specific AI technology; the latter denotes the societal dynamics elicited by its anticipated or actual emergence. Paradoxically, the very notion of "technological singularity" can become self-defeating, insofar as it mobilises countervailing action that alters or prevents the outcome it predicts.

Under such conditions, events can take different turns. In one version of the future, social singularity may curtail further AI advance and prevent technological singularity from occurring. Some existing AI-futures accounts already imagine backlash or conflict strong enough to halt or fragment progress (Yampolskiy 2015; de Garis 2005; Butler 1917), although they generally do not integrate these dynamics into a broader analysis of concurrent societal conditions. In another version, technological singularity may still follow, as warnings are ignored and AI pessimists suffer a Cassandra-like fate. Most scenarios, however, do not entertain the possibility of two partially independent singularities and instead treat them as a single event.

While the likelihood of social singularity, or some other form of strong social response should be taken seriously, its precise timing and character remain difficult to predict. Major social disruptions are notoriously hard to foresee given our limited knowledge of social processes[9] – not only for experts and the general public, but often even for the actors directly involved – as many decisive historical events have shown, including revolutions, coups d'état, and wars (Kuran 1989). This reflects the difficulty of disentangling the multiple factors that shape collective behaviour, as analysed in theories of threshold effects, social tipping, and self-organized criticality (Granovetter 1978; Zhukov et al. 2020; Wiedermann et al. 2020).

It is therefore plausible that major social conflicts or upheavals could emerge and fundamentally redirect future AI trajectories – yet their outcomes remain indeterminate. Such events may actually strengthen the existing *status quo* if they are suppressed, thereby delaying change for decades or permanently determining a society's path. Alternatively, they may produce sharp revolutionary turns directed against existing political power, dominant ideologies, or perceived enablers (i.e., AI companies and researchers in this case), or conversely counterrevolutionary turns directed against those attempting to restrain them. Because of the institutional divergences discussed earlier, these outcomes are unlikely to be uniform across societies and regions.

The social singularity can be precipitated by the fact that society as a whole may place even greater weight on the unpredictable hazards associated with AI ("unknown unknowns"). Long-term effects are often not visible at a technology's inception, when knowledge is limited and social uptake remains incomplete. Some risks can be anticipated and reduced to a tolerable level during deployment, others may become legible only much later, or not until they are deeply entrenched, as was the case with fossil fuels, combustion engines and climate change.

Societies have so far often avoided technological traps through self-correcting mechanisms that restored a workable equilibrium. Yet we should not succumb to survivorship bias. There is no assurance that humanity will be equally fortunate in the presence of cognitively superior AI, which could disrupt social homeostasis in unprecedented ways. Even peaceful coexistence would not dissolve the problem of control, which could still be lost permanently (Yampolskiy 2012, 2015). Awareness of this possibility already affects public sentiment and expert discourse and may, through multiple social mechanisms, feed back into decision-making, steer AI futures, and influence eventual outcomes.

## 3.3. Conclusion

Because of constraints of space, this paper has been limited to a selected set of mechanisms, and even those discussed here could not be elaborated in full detail. The central methodological claim, however, can now be stated as follows: plausible AI futures should be modelled as branching sociotechnical pathways rather than as technical trajectories with generic downstream social impacts. Many additional approaches from the social sciences and humanities could further enrich this analysis, including macroeconomic, cultural, anthropological, legal, and media-theoretical perspectives.

---

[9] It is fully conceivable, however, that unlike humans, a sufficiently advanced ASI, capable of large-scale data collection, and drawing on this particular knowledge, will be able not only to anticipate such events but could itself also trigger itself such disruptions or steer societies in that direction.

Moreover, although the paper identifies a range of social mechanisms likely to generate branching AI futures, it does not fully translate them into a systematic procedure for scenario construction. The framework proposed here should therefore be understood not as exhaustive, nor as a complete predictive model, but as a mechanism-based expansion of the scenario space beyond what gradual-disempowerment accounts typically consider. Future research could operationalise these mechanisms more systematically through comparative scenario methods and more explicit specification of the conditions under which particular mechanisms become causally decisive.

None of the foregoing precludes the possibility of AI takeover attempts, which may occur independently of public debate or social contestation. Multiple trajectories may unfold in parallel. Nor does this paper rule out low-probability, high-impact events: tipping points in innovation (e.g., breakthroughs in energy, or human biological enhancement), black-swan events in politics and society, or decisive idiographic factors that could rapidly alter AI trajectories. Timing also remains uncertain. The mechanisms analysed here may operate over years or decades; hence, it is fully imaginable that the future could unfold broadly along the lines described by gradual-disempowerment scenarios for an extended period, only to be interrupted by a sudden eruption of social unrest or revolution appearing seemingly "out of the blue" as has happened on a number of occasions in the past. The future naturally also depends on the degree to which alignment efforts succeed, whether in reality or in perception.

Even so, the mechanisms and historical analogies discussed in this paper suggest that AI futures are less likely to unfold in the smooth manner anticipated by gradual-disempowerment scenarios. More plausibly, they will be affected by branching sociotechnical dynamics in which framing, multi-speed adoption, and collective conflict all play constitutive roles.

# Funding

This work has been funded by a grant from the Programme Johannes Amos Comenius under the Ministry of Education, Youth and Sports of the Czech Republic, CZ.02.01.01/00/23_025/0008711.

# References

Abdelkarim S, Lu D, Flores D-L, Jaeggi S, Baldi P (2025) Evaluating the intelligence of large language models: a comparative study using verbal and visual IQ tests. Comput Hum Behav Artif Hum 5:100170. https://doi.org/10.1016/j.chbah.2025.100170

Acemoglu D (2024) The simple macroeconomics of AI. National Bureau of Economic Research (Working Paper 32487). https://doi.org/10.3386/w32487

Ainsworth MDS, Blehar MC, Waters E, Wall SN (2015) Patterns of attachment: a psychological study of the strange situation. Psychology Press, New York.

Amodei D (2024) Machines of Loving Grace: how AI could transform the world for the better. Dario Amodei. https://darioamodei.com/essay/machines-of-loving-grace. Accessed 17 March 2026.

Amodei D (2026) The adolescence of technology: confronting and overcoming the risks of powerful AI. Dario Amodei. https://www.darioamodei.com/essay/the-adolescence-of-technology#4-player-piano. Accessed 17 March 2026.

Anthis JR, Pauketat JVT, Ladak A, Manoli A (2025) Perceptions of sentient AI and other digital minds: evidence from the AI, Morality, and Sentience (AIMS) survey. In: Yamashita et al (eds) Proceedings of the 2025 CHI Conference on Human Factors in Computing Systems (CHI '25). Association for Computing Machinery, New York, NY, Article 10: 1-22. https://doi.org/10.1145/3706598.3713329

Anthropic (2024) Responsible scaling policy. Anthropic. https://www.anthropic.com/responsible-scaling-policy. Accessed 2 Mar 2026

Bacon A (2025) AI-powered children's toys are here, but are they safe? CNN Business, 1 December 2025. https://edition.cnn.com/2025/12/01/tech/ai-toys-safety. Accessed 6 March 2026.

Bandura A (1977) Social learning theory. Prentice-Hall, Englewood Cliffs

Bandura A, Barbaranelli C, Caprara GV, Pastorelli C (1996) Mechanisms of moral disengagement in the exercise of moral agency. J Pers Soc Psychol 71(2):364–374. https://doi.org/10.1037/0022-3514.71.2.364

Bengio Y (2023) FAQ on catastrophic AI risks. Yoshua Bengio. https://yoshuabengio.org/2023/06/24/faq-on-catastrophic-ai-risks/. Accessed 3 March 2026

Bengio Y et al (2026) International AI Safety Report 2026. Department for Science, Innovation and Technology (DSIT 2026/001), UK Government. https://internationalaisafetyreport.org/sites/default/files/2026-02/international-ai-safety-report-2026.pdf. Accessed 18 Feb 2026.

Biden J (2025) Remarks by President Biden in a Farewell Address to the Nation. The White House (archived), 15 January 2025. https://bidenwhitehouse.archives.gov/briefing-room/speeches-remarks/2025/01/15/remarks-by-president-biden-in-a-farewell-address-to-the-nation/. Accessed 17 Feb 2026.

Bostrom N (2014) Superintelligence: paths, dangers, strategies. Oxford University Press, Oxford

Bostrom N (2024) Deep Utopia: life and meaning in a solved world. Ideapress Publishing, Washington, DC

Bowlby J (1969) Attachment and loss, Vol. 1: Attachment. Penguin Books, New York

Brehm JW (1966) A theory of psychological reactance. Academic Press, New York

Bresnick S, Probasco ES, McFaul C (2026) China's AI arsenal: the PLA's tech strategy is working. Foreign Aff, 2 March 2026. https://www.foreignaffairs.com/china/chinas-artificial-intelligence-arsenal. Accessed 18 March 2026.

Burgoon M, Alvaro E, Grandpre J, Voulodakis M (2002) Revisiting the theory of psychological reactance: communicating threats to attitudinal freedom. In: Dillard JP, Pfau M (eds) The persuasion handbook: developments in theory and practice. SAGE, Thousand Oaks, pp 213–232

Bush V (1945) Science, the endless frontier: A report to the President on a program for postwar scientific research. United States Government Printing Office, Washington, DC.

Butler S (1917) Darwin among the machines. In: Jones HF (ed) The note-books of Samuel Butler: author of "Erewhon". E P Dutton & Co, New York.

Callon M (1998) The laws of the markets. Blackwell, Oxford.

Carlsmith J (2022) Is power-seeking AI an existential risk? arXiv preprint arXiv:2206.13353. https://doi.org/10.48550/arXiv.2206.13353

Cerullo M (2025) Barbie maker Mattel and OpenAI partner to develop AI-powered toys. CBS News, 12 June 2025. https://www.cbsnews.com/news/openai-mattel-barbie-artificial-intelligence-product/. Accessed 6 March 2026

Cheng M, Lee AY, Rapuano K et al (2026) Metaphors of AI indicate that people increasingly perceive AI as warm and human-like. Commun Psychol 4:8. https://doi.org/10.1038/s44271-025-00376-6

Cheng M, Yu S, Lee C, Khadpe P, Ibrahim L, Jurafsky D (2025) Social sycophancy: a broader understanding of LLM sycophancy. arXiv:2505.13995v1. https://doi.org/10.48550/arXiv.2505.13995. Available at: https://arxiv.org/html/2505.13995v1. Accessed 6 March 2026

Claburn T (2026) AI industry insiders launch site to poison the data that feeds them. The Register, 11 January 2026. https://www.theregister.com/2026/01/11/industry_insiders_seek_to_poison/. Accessed 17 March 2026.

CNN (2025) Anderson Cooper interviews Geoffrey Hinton on AI existential risk and "maternal instincts" proposal. Anderson Cooper 360° (transcript), CNN (13 August 2025). https://transcripts.cnn.com/show/acd/date/2025-08-13/segment/01. Accessed 18 February 2026.

Coeckelbergh M (2026) Technofascism: AI, Big Tech, and the new authoritarianism. AI Soc. https://doi.org/10.1007/s00146-026-02862-9

Cyberspace Administration of China et al (2023) Interim Measures for the Management of Generative Artificial Intelligence Services. English translation. China Law Translate. https://www.chinalawtranslate.com/en/generative-ai-interim-measures/ Accessed 3 March 2026

Dafoe A (2015) On technological determinism: A typology, scope conditions, and a mechanism. Sci Technol Human Values 40(6):1047–1076. https://doi.org/10.1177/0162243915579283

Dahrendorf R (1959) Class and class conflict in industrial society. Routledge & Kegan Paul, London

de Garis H (2005) The artilect war: cosmists vs. terrans: a bitter controversy concerning whether humanity should build godlike massively intelligent machines. Etc Publications, Palm Springs

DeepSeek (2025) DeepSeek-R1 release. DeepSeek. https://api-docs.deepseek.com/news/news250120. Accessed 2 Mar 2026.

Durand C (2024) How Silicon Valley unleashed techno-feudalism: the making of the digital economy. Verso, London

Eisenhower DD (1961) Farewell address to the nation. National Archives. https://www.archives.gov/milestone-documents/president-dwight-d-eisenhowers-farewell-address. Accessed 16 Feb 2026

Eisenstein EL (1980) The printing press as an agent of change: Communications and cultural transformations in early-modern Europe. Cambridge University Press, Cambridge

Epley N, Waytz A, Cacioppo JT (2007) On seeing human: a three-factor theory of anthropomorphism. Psychol Rev 114(4):864–886. https://doi.org/10.1037/0033-295X.114.4.864

European Union (2024) Regulation (EU) 2024/1689 of the European Parliament and of the Council laying down harmonised rules on artificial intelligence (Artificial Intelligence Act). Off J Eur Union. https://eur-lex.europa.eu/eli/reg/2024/1689/oj Accessed 2 Mar 2026

Forster K (2026) AI ‘arms race’ risks human extinction, warns top computing expert. Barron’s, 17 February 2026. https://www.barrons.com/news/ai-arms-race-risks-human-extinction-warns-top-computing-expert-74df6e59. Accessed 18 March 2026.

Geraci RM (2008) Apocalyptic AI: religion and the promise of artificial intelligence. J Am Acad Relig 76(1):138–166. https://doi.org/10.1093/jaarel/lfm101

Gilardi F, Kasirzadeh A, Bernstein A et al (2024) We need to understand the effect of narratives about generative AI. Nat Hum Behav 8:2251–2252. https://doi.org/10.1038/s41562-024-02026-z

Granovetter M (1978) Threshold models of collective behavior. Am J Sociol 83(6):1420–1443. http://www.jstor.org/stable/2778111 . Accessed 23 March 2026

Hafner J (2017) Alexa, are you turning my kid into a jerk? USA Today, 7 June 2017. https://eu.usatoday.com/story/tech/nation-now/2017/06/07/alexa-you-turning-my-kid-into-jerk/375949001/

Hall PA, Soskice D (eds) (2001) Varieties of capitalism: the institutional foundations of comparative advantage. Oxford University Press, Oxford. https://doi.org/10.1093/0199247757.001.0001. Accessed 27 Feb 2026.

Hedström P, Swedberg R (1996) Social mechanisms. Acta Sociol 39(3):281–308. https://doi.org/10.1177/000169939603900302

Herrman J (2023) What does it mean that Elon Musk's new AI chatbot is "anti-woke"? New York Magazine – Intelligencer, 7 November 2023. https://nymag.com/intelligencer/2023/11/elon-musks-grok-ai-bot-is-anti-woke-what-does-that-mean.html

Hobsbawm EJ, Rudé G (1969) Captain Swing. Lawrence and Wishart, London.

Horton D, Wohl RR (1956) Mass communication and para-social interaction: observations on intimacy at a distance. Psychiatry 19(3):215–229. https://doi.org/10.1080/00332747.1956.11023049

Huber L, Reynolds-Cuéllar P, DeVrio A, Raihan J, Sum CM, Dombrowski L, Zhang J, Becker CB, Irani L, Krafft PM, Hughes M (2025) From Tech Lash to Tech Fash: strategic reflections on a decade of collective organizing in computing. In: Adjunct proceedings of the sixth decennial Aarhus conference: Computing X Crisis (AAR Adjunct '25). Association for Computing Machinery, New York, NY, USA, Article 26, pp 1–4. https://doi.org/10.1145/3737609.3747097

Ishizaki R, Sugiyama M (2025) Large language models: assessment for singularity. AI Soc 40:5481–5491. https://doi.org/10.1007/s00146-025-02271-4

Jasanoff S, Kim S-H (eds) (2015) Dreamscapes of modernity: Sociotechnical imaginaries and the fabrication of power. University of Chicago Press, Chicago, IL. https://doi.org/10.7208/chicago/9780226276663.001.0001

Karnofsky H (2023) How we could stumble into AI catastrophe. Cold Takes. https://www.cold-takes.com/how-we-could-stumble-into-ai-catastrophe/. Accessed 16 Feb 2026

Kasirzadeh A (2025) Two types of AI existential risk: decisive and accumulative. *Philos Stud* 182:1975–2003. https://doi.org/10.1007/s11098-025-02301-3

Kelly G, Wallach E (2025) Hunger strikers in SF and London are calling for an end to AI. We Facetimed two of them. San Francisco Standard, 14 September 2025. https://sfstandard.com/2025/09/14/hunger-strike-ai-anthropic-google/. Accessed 2 Mar 2026

Khanal S, Zhang H, Taeihagh A (2025) Why and how is the power of Big Tech increasing in the policy process? The case of generative AI. Policy Soc 44(1):52–69. https://doi.org/10.1093/polsoc/puae012

Kilian KA (2026) Beyond accidents and misuse: decoding the structural risk dynamics of artificial intelligence. AI Soc 41:23–42. https://doi.org/10.1007/s00146-025-02419-2

Kleinert T, Waldschütz M, Blau J et al (2026) AI outperforms humans in establishing interpersonal closeness in emotionally engaging interactions, but only when labelled as human. Commun Psychol 4:23. https://doi.org/10.1038/s44271-025-00391-7

Kokotajlo D, Alexander S, Larsen T, Lifland E, Dean R (2025) AI 2027. AI Futures Project. https://ai-2027.com/ . Accessed 3 March 2026.

Krook J (2025) When autonomy breaks: the hidden existential risk of AI. AI Soc 40:6011–6024. https://doi.org/10.1007/s00146-025-02397-5

Kulveit J, Douglas R, Ammann N, Turan D, Krueger D, Duvenaud D (2025) Gradual disempowerment: systemic existential risks from incremental AI development. arXiv:2501.16946. https://arxiv.org/abs/2501.16946

Kuran T (1989) Sparks and prairie fires: a theory of unanticipated political revolution. Public Choice 61:41–74. https://doi.org/10.1007/BF00116762

Lee Y, Kim S-H (2025) Exploring dimensions of perceived anthropomorphism in conversational AI: implications for human identity threat and dehumanization. Comput Hum Behav Artif Hum 5:100192. https://doi.org/10.1016/j.chbah.2025.100192

Levin SA, Milner HV, Perrings C (2021) The dynamics of political polarization. Proc Natl Acad Sci U S A 118(50):e2116950118. https://doi.org/10.1073/pnas.2116950118

MacKenzie D (2006) An engine, not a camera: How financial models shape markets. MIT Press, Cambridge, MA.

Malfacini K (2025) The impacts of companion AI on human relationships: risks, benefits, and design considerations. AI Soc 40:5527-5540. https://doi.org/10.1007/s00146-025-02318-6

Manow P (2021) Models of the welfare state. In: Béland D, et al. (eds) The Oxford handbook of the welfare state, 2nd edn. Oxford University Press, Oxford. https://doi.org/10.1093/oxfordhb/9780198828389.013.45. Accessed 3 Mar 2026.

Marx K (1976) Capital: a critique of political economy, vol 1. Translated by Fowkes B. Penguin Books, Harmondsworth

Marx K, Engels F (2002) The Communist manifesto. Edited by Stedman Jones G. Penguin Books, London

Maslej N, Fattorini L, Perrault R et al (2025) Artificial intelligence index report 2025. Stanford University Human-Centered Artificial Intelligence, Stanford, CA. https://hai-production.s3.amazonaws.com/files/hai_ai_index_report_2025.pdf. Accessed 18 Feb 2026

McAdam D, Tarrow S, Tilly C (2001) Dynamics of contention. Cambridge University Press, Cambridge. https://doi.org/10.1017/CBO9780511805431

McCarthy JD, Zald MN (1977) Resource mobilization and social movements: a partial theory. Am J Sociol 82:1212–1241

Merton RK (1948) The self-fulfilling prophecy. Antioch Rev 8(2):193–210. https://doi.org/10.2307/4609267

Meta AI (2024) The Llama 3 herd of models. Meta. https://ai.meta.com/research/publications/the-llama-3-herd-of-models/. Accessed 2 Mar 2026.

Metcalf T (2025) AI safety and regulatory capture. AI Soc. https://doi.org/10.1007/s00146-025-02534-0

Mills CW (1959) The sociological imagination. Oxford University Press, New York

Naseh A, Chaudhari H, Roh J, Wu M, Oprea A, Houmansadr A (2025) R1dacted: Investigating local censorship in DeepSeek's R1 language model. arXiv. https://arxiv.org/abs/2505.12625 .

NIST (2024) Artificial intelligence risk management framework: generative artificial intelligence profile. National Institute of Standards and Technology, Gaithersburg. https://doi.org/10.6028/NIST.AI.600-1

Nozawa H, Vincze D, Sawada R, Niitsuma M, Kato M (2026) Child-Robot Bonding as a Safe Haven: Reducing Stress in Children and Mothers During Separation. ACM Trans Hum-Robot Interact. https://doi.org/10.1145/3796516

OECD (2025) AI openness: A primer for policymakers. OECD Artificial Intelligence Papers No. 44, Organisation for Economic Co-operation and Development, Paris. https://www.oecd.org/content/dam/oecd/en/publications/reports/2025/08/ai-openness_958d292b/02f73362-en.pdf. Accessed 2 Mar 2026.

Oltman S (2026) How the anxiety over AI could fuel a new workers' movement. The Guardian, 19 February 2026. https://www.theguardian.com/technology/ng-interactive/2026/feb/19/ai-work-future. Accessed 2 Mar 2026.

OpenAI (2023) GPT-4 technical report. https://doi.org/10.48550/arXiv.2303.08774

OpenAI (2023) Planning for AGI and beyond. OpenAI. https://openai.com/index/planning-for-agi-and-beyond/. Accessed 16 Feb 2026

OpenAI (2025) Preparedness framework. OpenAI. https://openai.com/index/updating-our-preparedness-framework/. Accessed 2 Mar 2026.

Ord T (2020) The precipice: Existential risk and the future of humanity. Bloomsbury, London

Patel D (2025) Ilya Sutskever - We're moving from the age of scaling to the age of research. Dwarkesh Patel, 25 November 2025. https://www.dwarkesh.com/p/ilya-sutskever-2 . Accessed 2 Mar 2026

Perrigo B (2024) Meta's AI Chief Yann LeCun on AGI, open-source, and AI risk. TIME. https://time.com/6694432/yann-lecun-meta-ai-interview/. Accessed 18 February 2026

Pew Research Center (2025) How the US public and AI experts view artificial intelligence. https://www.pewresearch.org/internet/2025/04/03/how-the-us-public-and-ai-experts-view-artificial-intelligence/. Accessed 18 Feb 2026

Pierson P (2000) Increasing returns, path dependence, and the study of politics. Am Polit Sci Rev 94:251–267. https://doi.org/10.2307/2586011

Puffert DJ (2009) Tracks across continents, paths through history: the economic dynamics of standardization in railway gauge. University of Chicago Press, Chicago

Qi T, Liu H, Huang Z (2025) An assistant or A friend? The role of parasocial relationship of human-computer interaction. Comput Hum Behav 167:108625. https://doi.org/10.1016/j.chb.2025.108625

Rickert H (1902) Die Grenzen der naturwissenschaftlichen Begriffsbildung: Eine logische Einleitung in die historischen Wissenschaften. Mohr, Tübingen

Roberts H, Hine E, Taddeo M, Floridi L (2024) Global AI governance: barriers and pathways forward. Int Aff 100(3):1275–1286. https://doi.org/10.1093/ia/iiae073

Roland A (2016) War and technology: A very short introduction. Oxford University Press, Oxford

Romeo D. (2025) A 2032 Takeoff Story. LessWrong, 6 November 2025. https://www.lesswrong.com/posts/yHvzscCiS7KbPkSzf/a-2032-takeoff-story. Accessed 20 March 2026.

Shang H, Liu X (2025) Mutual wanting in human-AI interaction: empirical evidence from large-scale analysis of GPT model transitions. arXiv:2510.24796. https://arxiv.org/abs/2510.24796

Şimşek C, Yaşar AG (2025) From rejection to regulation: mapping the landscape of AI resistance. SSRN working paper. Project page: Sciences Po Digital, Governance and Sovereignty Chair. https://www.sciencespo.fr/public/chaire-numerique/wp-content/uploads/2025/05/compressed-Simsek-and-Yasar-AI-Resistance-Report-publication-ready-2.pdf

Singler B (2024) Religion and artificial intelligence: an introduction. Routledge, Abingdon. https://doi.org/10.4324/9781003256113

Skjuve M, Følstad A, Fostervold KI, Brandtzaeg PB (2021) My Chatbot Companion – a study of human-chatbot relationships. Int J Hum Comput Stud 149:102601. https://doi.org/10.1016/j.ijhcs.2021.102601

Stanford Institute for Human-Centered Artificial Intelligence (2025) AI index report 2025. Stanford University, Stanford. https://hai.stanford.edu/ai-index/2025-ai-index-report. Accessed 2 Mar 2026.

Tajfel H, Turner JC (1979) An integrative theory of intergroup conflict. In: Austin WG, Worchel S (eds) The social psychology of intergroup relations. Brooks/Cole, Monterey, pp 33–47

Tarrow S (1998) Power in movement: social movements and contentious politics, 2nd edn. Cambridge University Press, Cambridge

The White House (2025) Winning the AI Race: America's AI Action Plan. The White House, Washington, DC. https://www.whitehouse.gov/wp-content/uploads/2025/07/Americas-AI-Action-Plan.pdf. Accessed 17 Feb 2026.

Thomas WI, Thomas DS (1928) The child in America: Behavior problems and programs. Knopf, New York

Tilly C (1978) From mobilization to revolution. McGraw-Hill, New York

Ulam S (1958) Tribute to John von Neumann. Bull Am Math Soc 64(3, Part 2):1–49. https://doi.org/10.1090/S0002-9904-1958-10189-5

Varoufakis Y (2023) Technofeudalism: what killed capitalism. The Bodley Head, London

Weber M (1978) Economy and society: An outline of interpretive sociology. University of California Press, Berkeley

Williamson OE (1985) The economic institutions of capitalism: firms, markets, relational contracting. Free Press, New York

Wiedermann M, Smith EK, Heitzig J et al (2020) A network-based microfoundation of Granovetter's threshold model for social tipping. Sci Rep 10:11202. https://doi.org/10.1038/s41598-020-67102-6

Winner L (1977) Autonomous technology: Technics-out-of-control as a theme in political thought. MIT Press, Cambridge, MA

Writers Guild of America (2023) Summary of the 2023 WGA MBA: AI provisions (contract summary). Writers Guild of America, Los Angeles. https://www.wga.org/contracts/contracts/mba/summary-of-the-2023-wga-mba. Accessed 2 Mar 2026.

Yampolskiy RV (2012) Leakproofing the singularity: Artificial intelligence confinement problem. J Conscious Stud 19(1–2):194–214

Yampolskiy RV (2015) Artificial superintelligence: a futuristic approach. Chapman and Hall/CRC, Boca Raton. https://doi.org/10.1201/b18612

YouGov (2025) Americans are increasingly likely to say AI will negatively affect society. YouGov, 18 July 2025. https://today.yougov.com/politics/articles/52615-americans-increasingly-likely-say-ai-artificial-intelligence-negatively-affect-society-poll. Accessed 16 Feb 2026

YouGov (2026) Most Americans say AI will reduce the number of jobs in the U.S. https://today.yougov.com/technology/articles/54123-most-americans-say-ai-artificial-intelligence-will-reduce-number-jobs-in-us-united-states-february-13-16-2026-economist-yougov-poll. Accessed 18 Feb 2026

Yudkowsky E (2008) Artificial intelligence as a positive and negative factor in global risk. In: Bostrom N, Ćirković MM (eds) Global catastrophic risks. Oxford University Press, Oxford, pp 308–345. https://doi.org/10.1093/oso/9780198570509.003.0021

Zhou V (2025) AI is reshaping childhood in China. Rest of World, 1 October 2025. https://restofworld.org/2025/ai-china-childhood/. Accessed 6 March 2026

Zhukov D, Kunavin K, Lyamin S (2020) Online rebellion: self-organized criticality of contemporary protest movements. SAGE Open 10(2). https://doi.org/10.1177/2158244020923354

Znaniecki F (1934) The method of sociology. Farrar & Rinehart, New York.